\documentclass[twocolumn]{aastex62}

\graphicspath{{./}{figures/}}

\submitjournal{ApJL}

\usepackage{ulem}

\shorttitle{Type I outbursts in low eccentricity Be/X-ray binaries}
\shortauthors{A. Franchini \& R. G. Martin}

\begin{document}

\title{
Type I outbursts in low eccentricity Be/X-ray binaries}

\correspondingauthor{Alessia Franchini}
\email{alessia.franchini@unlv.edu}

\author{Alessia Franchini}
\affil{Department of Physics and Astronomy, University of Nevada
4505 South Maryland Parkway, Las Vegas, NV 89154, USA}

\author{Rebecca G. Martin}
\affiliation{Department of Physics and Astronomy, University of Nevada
4505 South Maryland Parkway, Las Vegas, NV 89154, USA}



\begin{abstract}

Type I outbursts in Be/X--ray binaries are usually associated with the eccentricity of the binary orbit. The neutron star accretes gas from the outer parts of the decretion disk around the Be~star at each periastron passage. However, this mechanism cannot explain type~I outbursts that have been observed in nearly circular orbit Be/X-ray binaries.
With hydrodynamical simulations and analytic estimates we find that in a circular orbit binary, a nearly coplanar disk around the Be~star can become eccentric. The extreme mass ratio of the binary leads to the presence of the 3:1 Lindblad resonance inside the  Be star disk and this drives eccentricity growth.
Therefore the neutron star can capture material each time it approaches the disk apastron, on a timescale up to a few percent longer than the orbital period.  
We have found a new application of this mechanism that is able to explain the observed type I outbursts in low eccentricity Be/X-ray binaries.

\end{abstract}


\section{Introduction} 
\label{sec:intro}

Be/X-ray binaries  are  composed typically of a Be star and a neutron star companion.
The Be star is surrounded by a geometrically thin Keplerian decretion disk \citep{Pringle1991,Lee1991,Porter1996,Cassinelli2002,Porter2003} that is truncated by the tidal force exerted by the companion \citep{Okazaki2002,Hayasaki2004,Martin2011}. These systems can be persistent sources of X-rays or they can be transient, meaning that they alternate quiescent states with outbursts.
The outbursts are divided into two categories: type I and type II (or giant) depending on the duration and on the peak X-ray luminosity \citep{Stella1986,Negueruela1998,Okazaki2002}. Type I occur on the orbital period and are usually less bright ($L_{\rm X}=0.01-0.1\,L_{\rm Edd}$) while type II last several orbital periods and are brighter ($L_{\rm X}>0.1\,L_{\rm Edd}$).

The majority of the observed systems have eccentric orbits.
This naturally explains the occurrance of type~I outbursts when the neutron star passes through periastron and collects material from the outer parts of the Be star decretion disk \citep{Okazaki2001,Negueruela2001}.
However there are some examples of Be/X-ray binaries with nearly circular orbits that show type I outbursts \citep{Pfahl2002,Reig2007}. For example, XTE J1948+32, which has an orbital eccentricity of $e=0.03$ and orbital period $40.4$ day  \citep{Raguzova2005,Reig2007,Reig2013}.
 GS 0834-430 has orbital eccentricity $e=0.12$ and a $105.8$ days orbital period and showed five type~I outbursts reoccurring on a timescale of $107\,\rm day$ \citep{Wilson1997,Townsend2011,Cheng2014}. XTE J1543-568 has orbital eccentricity  $e<0.03$ and a $75.6$ days orbital period \citep{Reig2007,Reig2013,Cheng2014}. Finally, 2S~1553--542 has an obital eccentricity $e<0.09$ and orbital period $30.6\,\rm day$ \citep{Reig2007}.

In this Letter, we focus on the mechanism that leads to type~I  outbursts in nearly circular Be/X-ray binaries. In Section~\ref{sec:sim} we describe the results of hydrodynamical simulations of a low eccentricity Be/X-ray binary system. We find that the disk becomes eccentric and this leads to type~I like outbursts when the neutron star is close to the disk apastron.   In Section~\ref{sec:analytic} we explain the disk eccentricity growth by the presence of the 3:1 Lindblad resonance within the disk  \citep{Lubow1991a,lubow1991b,Lubow1992}. Previous studies suggested that this mechanism could not drive eccentricity growth in Be star discs since the disc may be truncated at the 3:1 resonance location \citep[e.g.][]{Okazaki2001}. However, we show that the disc is able to extend farther out than the 3:1 resonance. We draw our conclusions in Section~\ref{sec:disc}.

\section{Hydrodynamical simulations} 
\label{sec:sim}

We use the Smoothed Particle Hydrodynamics (SPH) code {\sc phantom} \citep{Lodato2010,Price2017} to model a binary system composed of a Be star surrounded by a coplanar accretion disk and a companion neutron star on a circular orbit.   Disks in binary systems have been extensively studied with {\sc phantom}  \citep[e.g.][]{Nixon2012,Nixon2015,Franchini2019}. We perform SPH simulations with $N=5 \times 10^5$ particles. 
The resolution of the simulation depends on $N$, the viscosity parameter $\alpha$ and the disk scale height $H$. 
The \cite{SS1973} viscosity parameter is modelled by adapting artificial viscosity according to the approach of \cite{Lodato2010}. With this choice of parameters the disc is initially resolved with a shell-averaged smoothing length per scale height of $\left< h\right>/H=0.628$ and this does not change much over the simulation.

The binary components are modelled as sink particles with masses $M_*=18\,M_{\odot}$ and $M_{\rm NS}=1.4\,M_{\odot}$. The binary mass ratio is therefore $q=M_{\rm NS}/M_\star=0.078$.
We choose a binary separation of $a=95\,R_{\odot}$, corresponding to an orbital period of $P_{\rm b}=24.3$ days. The orbital period is $P_{\rm b} = 2\pi/\Omega_{\rm b}$, where $\Omega_{\rm b}=\sqrt{G(M_\star+M_{\rm NS})/a^3}$.  We discuss how this choice of binary orbital period affects our results in Section~\ref{orbp}.
The accretion radii of the sink particles are chosen to be $R_{\rm acc,\star}=8\,R_{\odot}$ and $R_{\rm acc,NS}=0.5\,R_{\odot}$  for the Be star and the neutron star, respectively. Particles inside these radii are accreted onto the respective sink particle.
The accretion disk around the Be star extends from $R_{\rm in}=8\,R_{\odot}$ to an outer radius $R_{\rm out}=50\,R_{\odot}$ initially, has a total initial mass of $M_{\rm d}=1\times10^{-8}M_{\odot}$. We take the viscosity parameter to be $\alpha=0.3$, which is typical for fully ionized accretion disks \citep{King2007,Rimulo2018,Martin2019b}. We assume a globally isothermal equation of state for the gas with $H/R=0.01$ at the disk inner edge. This leads to $H/R=0.025$ at the initial disc outer radius. 
The initial surface density profile is $\Sigma \propto R^{-p}$ with $p=1$. The neutron star does not have an accretion disk at the beginning of the simulation.
The Be star accretion disk initially expands slightly, reaching the tidal truncation radius \citep{Artymowicz1994,Paczynski1977}.

\begin{figure*}
    \includegraphics[width=\textwidth]{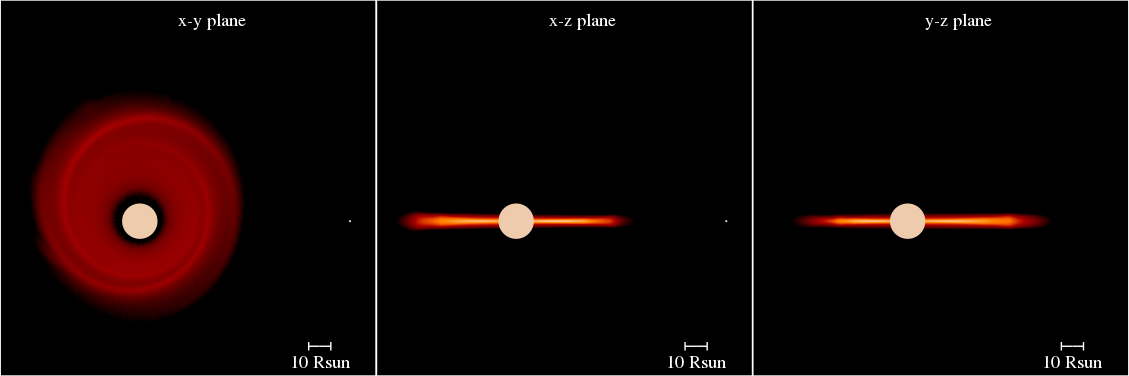}
    \caption{Column density of the Be star accretion disk from the SPH simulation at time $t=30\,P_{\rm b}$. The Be star is represented by the large white circle while the small white circle represents the companion neutron star. The size of the circle denotes the accretion radius of the sink. The left panel shows the view looking down on the $x-y$ binary orbital plane while the middle and right panels show the view in the $x-z$ and $y-z$ planes, respectively. The disk is initially circular and coplanar to the binary orbital plane. The Be star disk becomes eccentric due to the presence of the 3:1 resonance within the disk. }
    \label{fig:sph}
\end{figure*}

\begin{figure}
    \centering
    \includegraphics[width=\columnwidth]{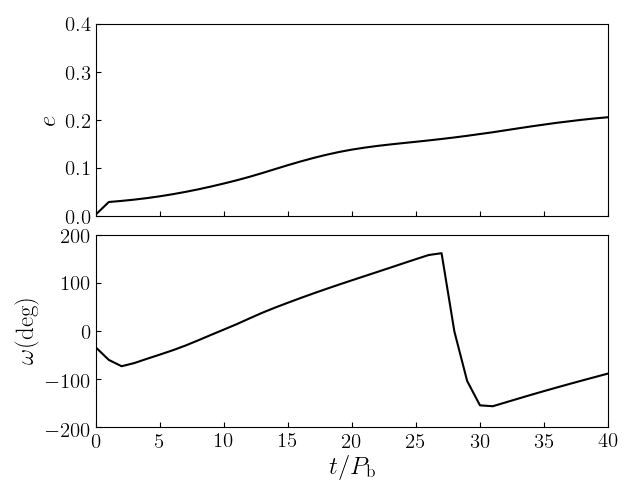}
    \caption{Density weighted average disk eccentricity (upper panel) and argument of periapsis (lower panel) evolution of the disk around the Be star. The Be star disk starts with zero eccentricity and coplanar to the binary plane.  
    }
    \label{fig:paramsvst}
\end{figure}

\begin{figure}
    \centering
    \includegraphics[width=\columnwidth]{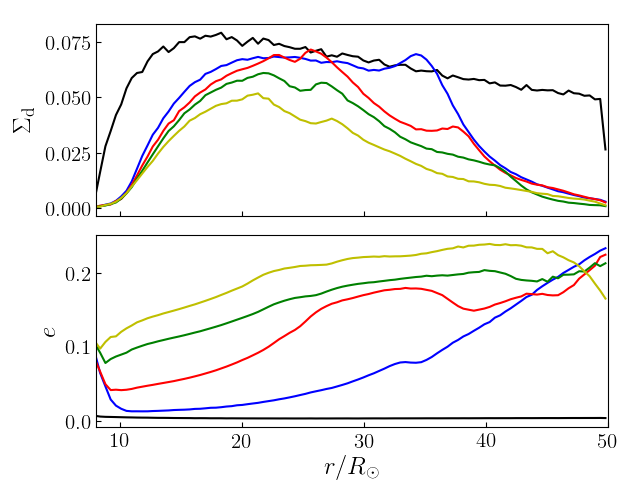}
    \caption{ Surface density profile (upper panel) and eccentricity (lower panel) of the Be star disk as a function of radius. The  lines represent the quantities at time $t=0$ (black), $t=10\,P_{\rm b}$ (blue), $t=20\,P_{\rm b}$ (red), $t=30\,P_{\rm b}$ (green) and $t=40\,P_{\rm b}$ (yellow). 
    }
    \label{fig:paramsvsr}
\end{figure}

 Fig.~\ref{fig:sph} shows the disk column density at a time of $30\,P_{\rm b}$ in our simulation. The left, middle and right panels show the view in the $x-y$, $x-z$ and $y-z$ planes respectively. We can clearly see that the disk remains coplanar to the binary orbital plane but becomes eccentric. 

Fig. \ref{fig:paramsvst} shows the eccentricity and argument of periapsis angle evolution of the Be star disk.  Both quantities are measured from our simulation through a density weighted average over the radial extent of the disk.
The eccentricity vector of the disc is precessing around the Be star on a timescale of about $40 \,P_{\rm b}$ while the magnitude of the eccentricity grows. Figure \ref{fig:paramsvsr} shows the disk surface density (upper panel) and eccentricity (lower panel) profile at five different times $t=0,\,10,\,20,\,30,\,40\,P_{\rm b}$. 
Most of the material that is lost from the disc is accreted on to the Be star. A small fraction is transferred on to the neutron star and some forms circumbinary material \citep{Franchini2019}.
The bottom panel of Figure \ref{fig:paramsvsr} shows that the eccentricity growth begins in the outer parts of the disc and is communicated through pressure in the disc to the inner parts. 

\begin{figure}
    \centering
    \includegraphics[width=\columnwidth]{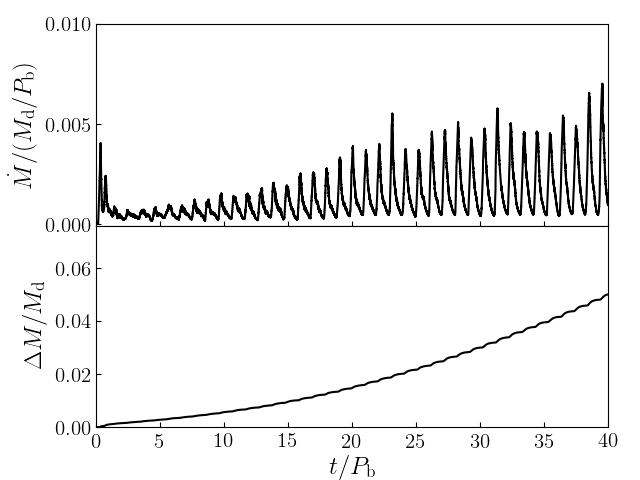}
    \caption{Upper panel: accretion rate onto the neutron star (at an accretion radius of $0.5\,R_{\odot}$). Bottom panel: total accreted mass onto the neutron star in units of the initial disk mass $M_{\rm d}=10^{-8}\,M_{\odot}$ vs time in units of the binary orbital period.     } 
    \label{fig:typeI-II}
\end{figure}

Figure \ref{fig:typeI-II} shows the accretion rate (upper panel) and the total mass accreted (lower panel) onto the neutron star as a function of time.
We see the occurrence of type~I outbursts in the accretion rate on a timescale of about $1.02\,P_{\rm b}$ starting after the first few binary orbits when the disk eccentricity starts to grow and the neutron star is able to capture material when it passes close to the disk apastron.

 We can estimate the X-ray luminosity reached during the outbursts to be $L_{\rm X}\approx 0.09\,L_{\rm Edd}$ (see Eq. 3 in \cite{Martin2014be}) which is the typical value observed in type~I outbursts.
Therefore for the accretion rate onto the neutron star to produce the observed X-ray luminosity we require the disc mass to be $M_{\rm d}\lesssim 10^{-8}\,M_{\odot}$.

\subsection{Effect of disk inclination}

We also investigated, through hydrodynamical simulations, the effect of the initial Be star disk inclination relative to the binary orbital plane on the disk eccentricity growth. 
We found that the eccentricity growth is not sufficient to induce outbursts if the inclination angle is above about $20^{\circ}$. Therefore the mechanism presented here would not drive type I outbursts in sources with large Be star spin misalignments with respect to the binary orbital plane. 

While we do not have observational evidence of the alignment of the disk, we expect the disk in a low eccentricity system to be close to coplanar to the binary orbit.  Prior to the formation of the neutron star, we expect that the spin of the Be star will be aligned to a circular binary orbit. Supernova explosions may be asymmetric leading to a kick on the newly formed neutron star \citep{Sutantyo1978}. This kick leads to an eccentric and inclined orbit \citep{Brandt1995,Martin2009}. Since the Be star binaries we consider in this work have low eccentricity, we also expect that they will have low misalignment angles \citep[e.g.][]{Podsiadlowski2004}.   

\section{Eccentricity growth in the Be star disk}
\label{sec:analytic}

The radius of the 3:1 resonance is 
\begin{equation}
R_{\rm res}=3^{-2/3}(1+q)^{-1/3}a
\label{rres}
\end{equation}
\citep[e.g.][]{Goodchild2006}. For our parameters, this is at a radius of $44.5\,\rm R_\odot$.
This leads to eccentricity growth within the disk at a rate 
\begin{equation}
    \lambda \simeq 2.1\,q^2\,\Omega_{\rm b}\,\frac{R_{\rm res}}{W}
    \label{eq:eccrate}
\end{equation}
\citep{Lubow1992}, where $W$ is the disk radial extent. 
The growth rate predicted for our set of parameters is $\lambda \simeq 0.002\,P_{\rm b}^{-1}$. The rate at which the Be star disk eccentricity grows as shown in Fig.~\ref{fig:paramsvst} is about 0.2 in 40 binary orbits, or $\lambda=0.005\,P_{\rm b}^{-1}$, higher than the theoretical prediction from equation~(\ref{eq:eccrate}). This discrepancy is likely due to the fact that equation~(\ref{eq:eccrate}) does not depend on the gas sound speed and viscosity. Higher viscosities and smaller disc aspect ratios both lead to faster disk eccentricity growth \citep[e.g.][]{Kley2008}.

The disk can become eccentric enough to fill the Be star Roche lobe and overflow onto the neutron star forming a misaligned and initially eccentric accretion disk \citep{Martin2014be,Franchini2019}.
The minimum eccentricity required for the disk to fill the Be star Roche lobe in this configuration can be obtained comparing the radius of the disc apastron with the Roche lobe radius \citep{Eggleton1983} and for our choice of parameters is $e_{\rm min}=0.14$ assuming an outer disc radius of $50\,R_{\odot}$. The accretion rate in Fig.~\ref{fig:typeI-II} shows that the outbursts become significant after about $10\,P_{\rm b}$ when the disk outer edge reaches an eccentricity of $0.2$ (see the blue line in the lower panel of Fig. \ref{fig:paramsvsr}).

The eccentric disk precesses in a prograde direction with period $P_{\rm p}$, which is much longer than the orbital period. Therefore, the companion neutron star reaches the line of apsides on a period slightly longer than the orbital period of the binary. The outburst period, $P_{\rm burst}$, is known as the apsidal superhump period in cataclysmic variables \citep[e.g.][]{Whitehurst1988,Whitehurst1991,Hirose1990} and it is related to the orbital period with
\begin{equation}
    P_{\rm burst} \simeq P_{\rm b}\left(1+\frac{P_{\rm b}}{P_{\rm p}}\right)
\end{equation}
\citep[e.g.][]{Murray1998} if the precession period is long compared to the binary orbital period.
The precession period for the disk can be approximated with
\begin{equation}
    P_{\rm p}=\frac{2\pi}{\omega_{\rm dyn}},
\end{equation}
where the dynamical component provides a prograde precession with frequency given by
\begin{equation}
    \omega_{\rm dyn}=\frac{1}{4r^2}\frac{d}{dr}\left(r^2\frac{db_{1/2}^{(0)}}{dr}\right)\frac{q}{\sqrt{1+q}}\Omega_{\rm b},
    \label{eq:omegadyn}
\end{equation}
where the radius is scaled to the binary orbital semi--major axis, $r=R/a$ and $b_{1/2}^{(0)}$ is the Laplace coefficient from celestial mechanics calculations \citep{Murray2000}. For a mass ratio $q=0.078$ and $r=0.47$ we find $\omega_{\rm dyn}/\Omega_{\rm b}=0.029$ and this corresponds to a precession period $P_{\rm p}=34.8\,P_{\rm b}$. The corresponding outburst period is $P_{\rm burst}=1.028\,P_{\rm b}$. 

This estimate is  an upper limit to the outburst period since the effects of pressure introduce a retrograde component \citep{Lubow1992,Murray1998,Murray2000,Kley2008}. The timescale between outbursts estimated from the simulation (see Fig.~\ref{fig:typeI-II}) is indeed slightly shorter than the one predicted here without considering pressure effects. For a sufficiently large disk aspect ratio, the precession may even become retrograde \citep[e.g.][]{Kley2008}. The precession rate is insensitive to the disk viscosity (unless the viscosity is very small), therefore accurate measurements of the type~I outburst period in low eccentricity X-ray binaries may also help to constrain the observed value for $H/R$.


\subsection{Effect of orbital period}
\label{orbp}

The simulation we presented in Section~\ref{sec:sim} is for a specific orbital period, but the mechanism can operate for a wide range of parameters provided that the disk is large enough to reach the 3:1 resonance location and the disc aspect ratio there is small enough.

For longer period Be/X-ray binaries the possibility of having type I outbursts in low eccentricity  systems depends on the Be star disk aspect ratio at the resonance radius. 
The strength of the 3:1 resonance scales with $(H/R)^{-2}$ \citep{Goodchild2006}. Therefore the larger the disk aspect ratio at the resonance radius, the weaker the resonance. 
For a fixed binary mass ratio, the resonance location and the tidal truncation radius both scale with semi--major axis. 
If the disk aspect ratio at the resonance radius is too large, the source is unlikely to undergo type I outbursts. Instead, its behavior  would be similar to a persistent X-ray emitter rather than a transient. Thus, this mechanism is more likely to operate for smaller orbital period binaries ($P_{\rm b}\lesssim 150$ days), assuming that the disks are flared. 

The inferred period for type~I outbursts taking into account the apsidal precession of the disk might be able to explain also the difference between the binary orbital period of $105.8\pm0.4$ days and the observed outburst timescale of $107$ days for the source GS 0834-430 \citep{Wilson1997}. These outbursts reoccur on a timescale of $1.007-1.015\,P_{\rm b}$.

\section{Conclusions} 
\label{sec:disc}

We have investigated type~I outbursts in nearly circular, relatively short period Be/X-ray binaries.
The presence of the 3:1 Lindblad resonance within the Be star disk leads to significant eccentricity growth. We found a new application of this mechanism to explain this type of outburst if the Be star decretion disk is close to coplanar with the binary orbital plane.
The neutron star is able to capture material every time it passes  the disk apastron thus producing type~I outbursts. 
This mechanism for driving type~I outbursts in a circular orbit binary relies on two system properties. First, the disk must be large enough to reach the location of the 3:1 Lindblad resonance given in equation~(\ref{rres}) and this is valid for binary mass ratios $q \lesssim 0.33$ \citep{FKR2002}. Secondly, the disk aspect ratio must be small enough at the resonance radius in order for the 3:1 Lindblad resonance to be able to drive eccentricity growth in the disk. This means that the mechanism is more likely to operate for shorter orbital period binaries, if the discs are flared.

\section*{Acknowledgements}

We thank Daniel Price for providing the {\sc phantom} code for SPH
simulations and acknowledge the use of SPLASH \citep{Price2007} for
the rendering of the figures. We acknowledge support from NASA through grant NNX17AB96G. Computer support was provided by UNLV's
National Supercomputing Center.


\end{document}